%% file: sample-sigplan.tex
\documentclass[sigplan,10pt]{acmart}
\renewcommand\footnotetextcopyrightpermission[1]{}
\AtBeginDocument{%
  }

\usepackage{xspace}
\usepackage{xurl}
\usepackage{makecell}
\usepackage{tabularx}
\usepackage{multirow}
\usepackage{verbatim}

\usepackage{enumitem}
\usepackage{algorithm}
\usepackage{algpseudocode}
\usepackage{multicol}
\usepackage{tcolorbox}

\begin{document}

\renewcommand{\b}[1]{\textbf{#1}}
\renewcommand{\i}[1]{\textit{#1}}
\renewcommand{\c}[1]{\texttt{#1}}
\renewcommand{\u}[1]{\underline{#1}}
\newcommand{\x}{$\times$\xspace}
\newcommand{\sysname}{\textsc{PipeANN-Filter}\xspace}
\newcommand{\specfilter}{speculative filtering\xspace}
\newcommand{\Specfilter}{Speculative filtering\xspace}
\newcommand{\SpecFilter}{Speculative Filtering\xspace}
\newcommand{\modify}[1]{\color{black} #1 \color{black}}
\newcommand{\extend}[1]{\color{black} #1 \color{black}}

\newcommand{\parahead}[1]{\par\smallskip\noindent\textbf{\textit{#1}}\xspace}

\title{\LARGE{\sysname: An Efficient Filtered Vector Search System on SSD}}


\author{Hao Guo}
\affiliation{%
  \institution{Tsinghua University}
  \city{Beijing}
  \country{China}
}

\author{Jiwu Shu}
\affiliation{%
  \institution{Tsinghua University}
  \city{Beijing}
  \country{China}
}

\author{Youyou Lu}
\authornote{Youyou Lu is the corresponding author (\href{mailto:luyouyou@tsinghua.edu.cn}{luyouyou@tsinghua.edu.cn}).}

\affiliation{%
  \institution{Tsinghua University}
  \city{Beijing}
  \country{China}
}

\include{section/abstract}




\settopmatter{printfolios=true}
\maketitle
\pagestyle{plain}

\input{section/intro}
\input{section/background}
\input{section/design}
\input{section/impl}
\input{section/evaluation}
\input{section/related_work}
\input{section/conclusion}


\bibliographystyle{ACM-Reference-Format}
\bibliography{sample-base}

\end{document}

%% file: section/abstract.tex
\begin{abstract}
We propose \sysname, an efficient filtered vector search system on SSD.
Unlike existing systems that explore only valid vectors (i.e., those satisfying the attribute constraints) during search,
\sysname explores a \emph{superset} of valid vectors, and performs attribute verification after getting the top-$k$ closest result vectors.
This allows \sysname to leverage probabilistic data structures (e.g., Bloom filters) to identify the superset, trading off a small number of false-positive vector explorations for a massive reduction in SSD I/O for attribute reading.
Evaluations show that \sysname improves search latency and throughput compared to state-of-the-art systems.
\sysname is open-source at \url{https://github.com/thustorage/PipeANN}.
\end{abstract}

%% file: section/intro.tex
\section{Introduction}
\label{sec:intro}

Real-world datasets associate high-dimensional vectors with attributes. 
Vectors capture data semantics~\cite{iclr13word2vec,sigmod20any2vec,icml21clip}, while each attribute provides a factual property of variable size.
To query such datasets, systems use filtered Approximate Nearest Neighbor Search (\emph{filtered ANNS}), which finds the top-$k$ closest \emph{valid vectors} (i.e., those satisfying attribute constraints) to a query. 
It is widely used in applications like search~\cite{vldb2020adbv} and retrieval-augmented generation~\cite{nips20rag,kdd24rag}. 
For example, an e-commerce platform~\cite{vldb2020adbv,middleware18jdecommerce} searches for items that are semantically similar to a reference image, while restricting the results to a specific price range and brand.

Filtered ANNS explores vectors in a graph-based index~\cite{www23filtered,sigmod24acorn,sigmod24ung,vldb25unify}, where vectors represent nodes and are connected by edges. 
Attribute filtering can occur before ANNS by scanning attributes (\emph{pre-filtering}), or after ANNS by verifying the result vectors (\emph{post-filtering}).
The graph-based index also enables \emph{in-filtering}: when exploring a vector, the system filters its neighbors, in order to traverse only the sub-graph of valid vectors.
These three mechanisms are complementary; the optimal choice depends on the query selectivity, namely the ratio of valid vectors in the dataset~\cite{sigmod26fannsbenchmark}.

Traditional ANNS systems store all data in memory.
However, as datasets scale to billions of vectors, both academia~\cite{nips19diskann,nips21spann} and industry~\cite{sigmod21milvus,vldb2020adbv} manage vectors on Solid-State Drives (SSDs) for ANNS. 
Filtered ANNS datasets share the same scale: Alibaba~\cite{vldb2020adbv} manages 830 million vectors with 21 attributes each, and LAION400M~\cite{arxiv21laion400m} contains 400 million vectors with 15 attributes each.
Therefore, it is crucial to manage both vectors and attributes on SSDs for filtered ANNS; yet, we find this remains underexplored.

Unlike in-memory systems, filtered ANNS on SSD introduces I/O overheads, which make pre- and in-filtering inefficient.
For in-filtering, exploring a vector requires reading hundreds of neighbor vectors (for distance comparison) and their attributes (for filtering).
While in-memory compressed vectors can mitigate vector reads~\cite{nips19diskann}, attribute reads still incur massive random SSD I/O, limiting search throughput.
Similarly, pre-filtering requires on-SSD attribute scans, which are highly expensive, especially for multi-attribute constraints.
A trivial workaround is to store all attributes in memory. 
However, this sacrifices the memory efficiency of on-SSD ANNS. 
For instance, attributes in LAION400M take 60GB, far exceeding the 24GB used by compressed vectors.
This issue is more severe for datasets with richer attributes, requiring $\sim$300GB for 500 million e-commerce items~\cite{arxiv24blair}.

We find that the root cause of this I/O bottleneck is \emph{strict filtering}:
Both pre- and in-filtering require every vector explored during the search \emph{must be} valid.
Therefore, before exploring any vector, the system must fetch its exact attributes from the SSD, causing the I/O bottleneck.

We argue that strict filtering is unnecessary and propose \emph{\specfilter}.
Instead of filtering and exploring only valid vectors, \specfilter allows exploring a \emph{superset} of them (allowing false positives), and performs attribute verification after getting the closest result vectors.

\Specfilter enables a tradeoff: accepting false-positive explorations to reduce attribute reads. 
We demonstrate how this tradeoff boosts performance through two examples.
For in-filtering, \specfilter can filter neighbors using memory-efficient probabilistic data structures (e.g., Bloom filters for label membership checks), instead of reading their attributes from the SSD.
For pre-filtering, instead of scanning all involved attributes, \specfilter can prune constraints to generate the superset (e.g., executing only the most lightweight branch in a multi-attribute \c{AND} constraint). 
Attribute verification is then executed only for the result vectors of ANNS.

One might assume that reading exact attributes for verification introduces additional I/O overhead.
However, we find that it incurs little to no extra I/O:
On-SSD ANNS first estimates distances using in-memory compressed vectors, and then fetches the closest full-precision vectors from the SSD for re-ranking. 
These re-ranked vectors are exactly what \specfilter verifies. 
By co-locating a vector and its attributes, a single I/O retrieves both. 
If both share the same SSD page, reading attributes introduces no extra I/O.

Interestingly, we find that some false-positive explorations improve search accuracy. 
Since the base ANNS graph is sparse ($\sim$100 neighbors per node), low query selectivity often leaves nodes without valid neighbors, fragmenting the valid vectors into disconnected sub-graphs.
This traps traditional in-filtering, which explores only valid vectors, in local optima, resulting in low accuracy. 
By exploring invalid vectors, \specfilter bridges these disconnected sub-graphs, allowing the search to find more valid vectors.

Despite its potential, applying \specfilter to an on-SSD filtered ANNS system is challenging. 
First, cost estimation is required to route each query to its suitable filtering mechanism. 
This is complex because the model must consider both compute and I/O costs, while accounting for false positives. 
Second, the data structures for getting the superset must simultaneously achieve low SSD I/O, a low false-positive rate, and a small memory footprint.

To tackle these challenges, we design \sysname, an efficient filtered ANNS system on SSD. 
For cost estimation, \sysname computes the "equivalent" search parameters required to find a given number of valid vectors, based on the estimated false-positive rate and query selectivity. 
Both compute and I/O costs can be estimated using the search parameters~\cite{pr1980rng}. 
For efficient superset generation, \sysname achieves memory efficiency and low SSD I/O via a two-level data structure design. It combines in-memory Bloom filters with on-SSD inverted indexes for label constraints~\cite{www23filtered}, and combines in-memory quantized values with on-SSD sorted indexes for range constraints~\cite{sigmod24serf}.
Furthermore, \sysname accelerates Boolean combinations (\c{AND}, \c{OR}) by pruning heavy constraints, and provides a flexible interface to support custom, user-defined attribute constraints.

We evaluate \sysname to show its efficacy. 
On million-scale datasets targeting a 0.9 recall, \sysname reduces latency by at least 89.5\% and delivers at least 32.3$\times$ higher throughput compared to Milvus~\cite{sigmod21milvus}, a state-of-the-art vector database using pre-filtering. 
Compared to PipeANN-BaseFilter~\cite{osdi25pipeann}, an optimized baseline that dynamically routes queries to pre- or post-filtering, \sysname achieves at least 1.71$\times$ higher throughput with comparable latency. 
This performance boost scales to a 100-million-vector dataset. 
Furthermore, against Filtered-DiskANN~\cite{www23filtered}, a strict in-filtering system using customized graph structures, \sysname achieves 4.35$\times$ higher throughput and reduces latency by 77.6\%.
Notably, \sysname also achieves higher peak recall, as its false-positive explorations improve graph connectivity.

In summary, this paper makes the following contributions:
\begin{itemize}[itemsep=2pt,topsep=2pt,parsep=0pt,leftmargin=12pt]
    \item We analyze the I/O overheads of filtered ANNS on SSD, identifying strict filtering as the root cause (\S\ref{sec:background}).
    \item We argue that strict filtering is unnecessary and propose \specfilter, which allows exploring a superset of valid vectors. This enables a tradeoff between false-positive vector exploration and attribute-filtering I/O (\S\ref{sec:design}).
    \item We design \sysname, a filtered ANNS system on SSD. Leveraging \specfilter, \sysname supports arbitrary attribute constraints while reducing SSD I/O via memory-efficient probabilistic data structures (\S\ref{sec:impl}).
    \item We evaluate \sysname to demonstrate its state-of-the-art efficacy in on-SSD filtered ANNS (\S\ref{sec:evaluation}).
\end{itemize}

%% file: section/background.tex
\section{Background and Motivation}
\label{sec:background}

In this section, we first introduce on-SSD graph-based ANNS.
We then describe filtered ANNS mechanisms and analyze their overheads when deployed on the SSD.

\subsection{On-SSD Graph-based ANNS}
\label{sec:bg:ondisk}

Given a query vector, Approximate Nearest Neighbor Search (ANNS) finds its top-$k$ nearest \emph{base vectors} (vectors in the dataset).
The search is approximate for efficiency, meaning that it may only find a subset of the exact top-$k$ vectors.

\begin{figure}[t]
\begin{center}
\includegraphics[width=\linewidth]{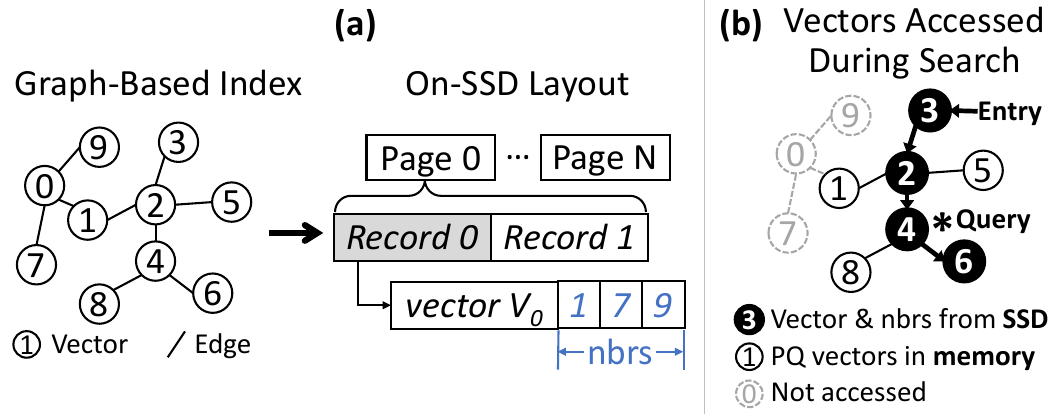}
\end{center}
\caption{\label{fig:ondisk} 
Overview of an on-SSD graph-based ANNS index. 
(a) On-SSD data layout. Each record stores a full-precision vector and its neighbor IDs. 
(b) Vector access pattern during a search. 
Records of vectors along the search path are fetched from the SSD. 
Their neighbors' PQ-compressed vectors are accessed in memory for distance comparison, without involving the SSD.
Other vectors are not accessed.
}
\end{figure}

\parahead{Index structure.}
As shown in Figure~\ref{fig:ondisk}(a), a graph-based ANNS index organizes base vectors into a directed graph, where each vector represents a node and is connected by edges.
On the SSD, this graph is stored as a set of adjacency lists.
We define each adjacency list as a \emph{record}, which contains a full-precision vector and its out-neighbor IDs (e.g., $V_0$ and its neighbors 1, 7, 9).
In memory, the index maintains compressed vectors using product quantization (PQ)~\cite{cvpr13pq} to reduce SSD reads during a search.
Typically, the memory-to-SSD space ratio is 1:10~\cite{nips19diskann}.

\parahead{Search algorithm.}
Graph-based ANNS follows a best-first search algorithm~\cite{tpami2018hnsw,vldb2019nsg,nips19diskann}.
As shown in Figure~\ref{fig:ondisk}(b), starting from an entry node (Node 3), the search navigates toward the query vector step by step.
In each step, the algorithm explores one vector in the \emph{candidate pool}: it fetches the current closest vector and its neighbor IDs from the SSD (in a single record).
Using the neighbor IDs, it retrieves the PQ-compressed neighbor vectors from memory and computes their approximate distances to the query, without involving the SSD.
Then, it adds the neighbors to the candidate pool, sorts them using PQ distance, and decides which vector to explore in the next step.
The search terminates upon reaching a local optimum, specifically when the top-$L$ vectors in the candidate pool can no longer be updated with closer neighbors.
The explored vectors are re-ranked using their full-precision distances to produce the results.

\subsection{Filtered ANNS}

Given a query vector, filtered ANNS finds the top-$k$ nearest base vectors that satisfy specific \emph{attribute constraints} (we call these \emph{valid} vectors).
In this scenario, a base vector is associated with several \emph{attributes} of various data types, such as integers, strings, or lists of labels.
An attribute constraint defines a condition an attribute must meet (e.g., falling within an $[l, r)$ range, matching a string prefix, or containing a specific label).
Multiple attribute constraints are connected using Boolean expressions (e.g., \c{AND}/\c{OR}).

\parahead{The \c{is\_member} abstraction.}
In this paper, we focus on filtered ANNS with \emph{arbitrary attribute constraints}.
We abstract these constraints into a single Boolean function: it takes the query's and a base vector's attributes as input, and outputs true if the vector is valid, or false otherwise.
Following existing systems~\cite{douze2024faiss,sigmod24acorn}, we call this function \c{is\_member}.

\parahead{Filtered ANNS mechanisms.}
There are three mechanisms for filtered ANNS: post-filtering, pre-filtering, and in-filtering.
Their efficiency varies with different \emph{query selectivity} (i.e., the ratio of valid vectors in the dataset)~\cite{sigmod26fannsbenchmark,vldb25unify}.

\begin{itemize}[itemsep=2pt,topsep=2pt,parsep=0pt,leftmargin=12pt]
    \item \emph{Post-filtering} first performs ANNS to get an initial vector set with more than $k$ vectors, and then filters the set to keep only the valid ones. 
    It is suitable for high query selectivity, where valid vectors are abundant.
    
    \item \emph{Pre-filtering} first scans the attributes to collect the valid vector set, and then performs a brute-force exact NNS on this set.
    It is suitable for low query selectivity, where post-filtering would require an exhaustively long graph traversal to find enough valid vectors.
    
    \item \emph{In-filtering} applies the filter dynamically during the ANNS process. 
    During graph traversal, it only explores valid vectors.
    It is suitable for moderate query selectivity: in this case, post-filtering still has high traversal overhead, while pre-filtering's brute-force NNS becomes expensive on the moderately sized subset.
\end{itemize}

\subsection{Analyzing Filtered ANNS on SSD}
\label{sec:moti:analyze}
\begin{figure}[t]
\begin{center}
\includegraphics[width=\linewidth]{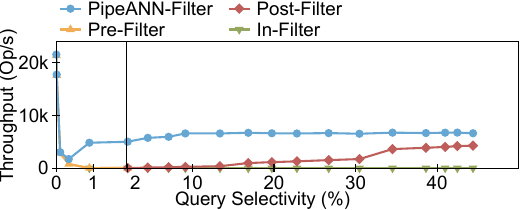}
\end{center}
\caption{\label{fig:sel-tput} Search throughput of different filtering mechanisms across varying selectivities.
Dataset: LAION100M. 
Target recall: 0.9.
The blue line shows the performance of our system, \sysname.
}
\end{figure}

In real-world vector datasets, vectors are usually associated with multiple attributes, which consume significant space.
For example, in the LAION400M dataset~\cite{arxiv21laion400m}, each vector has 15 attributes (e.g., image description, image width), making up 58.6 GB for 400 million vectors.
In Amazon's e-commerce dataset~\cite{arxiv24blair}, 500 million items contain $\sim$300GB attributes (e.g., product features and reviews).

Therefore, to support large-scale vectors with attributes, it is crucial to \emph{store both vectors and their attributes on SSDs} for filtered ANNS.
Unlike existing systems that store attributes in memory~\cite{www23filtered,vldb25unify}, checking whether a vector is valid requires reading its attributes from SSD, which incurs significantly higher I/O overheads.

In this paper, we explore \b{how to achieve efficient filtered ANNS on SSDs}.
To understand the I/O overheads of existing filtered ANNS mechanisms, we evaluate them on a 100M-scale dataset, LAION100M, using the range-filtering workload (detailed in \S\ref{sec:setup}). Figure~\ref{fig:sel-tput} shows the results: 

\parahead{Post-filtering works, but only for high selectivity.}
Post-filtering applies the filter to the initial ANNS results, which have already been fetched from the SSD during the search process.
By storing the attributes inside the same record as the vector, post-filtering introduces no separate I/O for attributes and thus remains highly efficient.
Consequently, Figure~\ref{fig:sel-tput} shows that it achieves reasonable throughput at high selectivities. 
However, when the selectivity drops below 10\%, its throughput drops to $<$300 QPS, as it should traverse a substantial part of the graph to find enough valid results.

\parahead{Pre-filtering is inefficient due to scan and compute.}
Pre-filtering requires an attribute scan across the dataset to collect valid vectors, followed by an exact NNS step.
As shown in Figure~\ref{fig:sel-tput}, pre-filtering only performs well at extremely low selectivities ($<$1\%). Once the selectivity reaches $\ge$1\% (requiring an attribute scan and NNS over $\ge$1 million vectors), its throughput degrades to $\le$100 QPS.
This is because, when the selectivity is not extremely low, the exact NNS becomes a CPU bottleneck. 
While this process can be I/O-efficient by comparing distances using PQ-compressed vectors in memory (and only fetching the top-$L$ vectors from the SSD for re-ranking), computing these distances for millions of compressed vectors remains compute-intensive.

Even when the selectivity is extremely low, the attribute scan may still be a bottleneck.
If the query involves multiple attributes or cannot be accelerated by attribute indexes, the system is forced to perform expensive attribute scans on the SSD just to identify the valid vectors.

\parahead{In-filtering is inefficient due to random SSD I/O.}
Recall the access pattern in Figure~\ref{fig:ondisk}(b): after fetching Record 2, the in-filtering mechanism needs to check if its neighbors (1, 4, 5) are valid, which requires reading their attributes from the SSD.
Because each vector typically has 32--128 neighbors that are randomly scattered across the graph~\cite{sigmod24starling}, reading their attributes translates to massive random SSD reads, which severely degrades search throughput.
Consistent with this analysis, Figure~\ref{fig:sel-tput} shows that in-filtering's throughput remains consistently $<$50 QPS, regardless of query selectivity.

\parahead{Root cause of I/O overheads: Strict filtering.}
In conclusion, to support arbitrary query selectivity, we must rethink the filtering mechanisms to eliminate these massive I/O overheads.
We find that these bottlenecks are fundamental to how current mechanisms operate. 
Specifically, both pre- and in-filtering enforce \emph{strict filtering}: they guarantee that every vector explored during the search \emph{must be} a valid vector. 
To maintain this strict guarantee, the algorithm must fetch and check the actual attributes from the SSD for every encountered vector, inevitably causing the I/O bottleneck.

%% file: section/design.tex
\section{\SpecFilter}
\label{sec:design}

We argue that strict filtering is not necessary and propose \emph{\specfilter}.
Instead of exploring only the valid vectors, \specfilter explores a \emph{superset} of them during pre- and in-filtering, and performs exact verification of attributes after getting the closest result vectors.

\parahead{Benefits.}
This design enables a new I/O tradeoff: we trade a slight increase in false-positive explorations for a massive reduction in attribute filtering I/O.
We find three key opportunities where this tradeoff boosts search efficiency:

First, to get this superset, we can significantly reduce SSD I/O by employing memory-efficient probabilistic data structures.
For example, a Bloom filter can quickly check if a base vector contains the queried tags.
This allows us to identify the superset entirely in memory, introducing only a low false-positive rate while bypassing the SSD.

Second, the exact verification can be piggybacked on the final re-ranking phase, often incurring little to no extra I/O.
Recall that on-SSD ANNS already fetches full-precision vectors for re-ranking.
We pack the attributes into the same record as the vector, and fetch them together during reranking.
This is effective because records are typically not page-aligned; their final page often contains significant unused space.
When the attributes are placed in this leftover space, fetching them does not incur extra I/O.

Third, during in-filtering, some false-positive explorations are actually useful; they can be repurposed to improve graph connectivity.
When query selectivity is moderately low, strict in-filtering prunes many edges, which breaks the connectivity and thus causes early search termination.
In \specfilter, the false positives naturally act as "bridge" nodes. 
Even though these nodes are invalid, exploring their neighbors helps reconnect the graph, thus reducing the chance for the search algorithm to fall into local optima.

\parahead{The \c{is\_member\_approx} abstraction.}
\Specfilter uses a new function, \c{is\_member\_approx}, to get the superset.
Unlike the original \c{is\_member} function, which inputs the base vector's attributes on SSD, \c{is\_member\_approx} inputs only the query's attributes and the base vector's ID.
It guarantees no false negatives: if it returns \c{false}, the vector is definitely invalid. 
If it returns \c{true}, the vector \emph{might} be valid (it belongs to the superset).
The original \c{is\_member} function is used in the final step for exact verification.

\begin{figure*}[t]
\begin{center}
\includegraphics[width=0.83\linewidth]{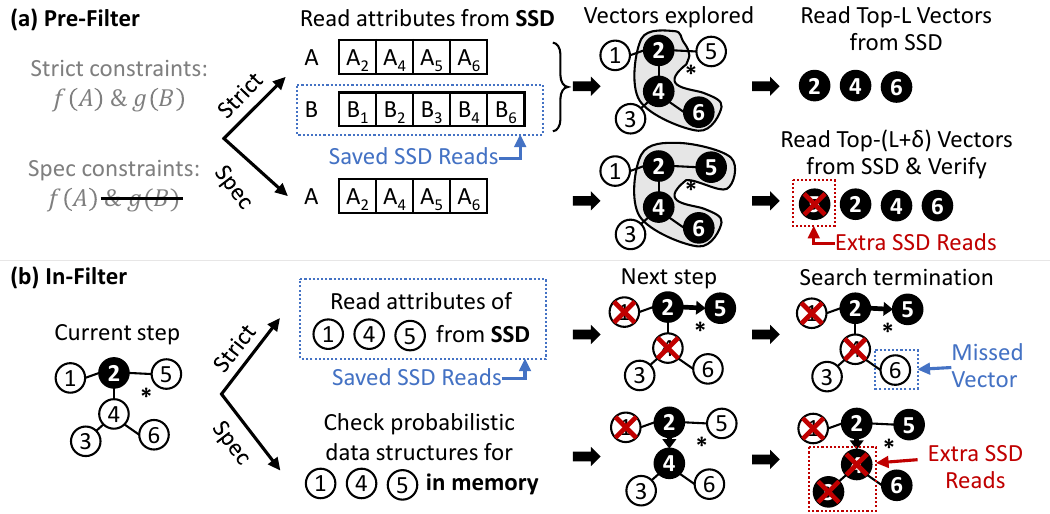}
\end{center}
\caption{\label{fig:spec-filter-overview} Comparison of speculative pre-/in-filtering with strict pre-/in-filtering.
}
\end{figure*}

\parahead{\Specfilter for pre-filtering and in-filtering.}
Leveraging \specfilter, we modify the execution flow of pre-filtering and in-filtering to reduce SSD reads.

\emph{Pre-filtering.} As shown in Figure~\ref{fig:spec-filter-overview}(a), strict pre-filtering scans all attribute indexes on the SSD to evaluate every constraint (e.g., $f(A)$ and $g(B)$). 
Instead, \specfilter gets a superset by evaluating only a partial set of constraints (e.g., $f(A)$). 
This reduces SSD reads by completely bypassing the index for $B$.
Next, it performs a brute-force NNS using PQ distances in memory to find the top-$(L+\delta)$ vectors. 
Finally, it fetches these vectors from the SSD for reranking and uses \c{is\_member} for exact verification. 
Fetching the $\delta$ extra vectors incurs additional SSD reads.

\emph{In-filtering.} Figure~\ref{fig:spec-filter-overview}(b) illustrates how \specfilter operates during the graph search. 
In a search step, strict in-filtering reads the attributes of all neighboring nodes (1, 4, and 5) from the SSD. 
Instead, \specfilter checks memory-efficient probabilistic data structures to filter out invalid neighbors. 
This eliminates the SSD reads for neighbor attributes. 
Since the check is probabilistic, the search might explore and verify some invalid nodes (like node 4), incurring extra reads from the SSD. 
However, some invalid nodes act as valuable "bridge" nodes. 
For instance, exploring node 4 enables the search algorithm to reach the valid node 6. 
This reduces the chance of early search termination and discovers vectors that strict filtering would otherwise miss.

\parahead{Unifying existing methods.}
Interestingly, \specfilter unifies existing filtering methods, which can be viewed as two special cases:
\begin{itemize}[itemsep=2pt,topsep=2pt,parsep=0pt,leftmargin=12pt]
    \item \emph{Strict pre- and in-filtering} are the strict extreme. Their \c{is\_member\_approx} is identical to \c{is\_member} with a zero false-positive rate.
    This pushes all the filtering work into the graph traversal step and needs no verification.
    \item \emph{Post-filtering} is the loose extreme. Its \c{is\_member\_approx} is a dummy function that always returns \c{true} with a maximum (1 - selectivity) false-positive rate.
    This pushes all the filtering work to the final verification.
\end{itemize}

\Specfilter enables new trade-offs between these two extremes.

%% file: section/impl.tex
\section{\sysname Design and Implementation}
\label{sec:impl}

\begin{figure}[t]
\begin{center}
\includegraphics[width=\linewidth]{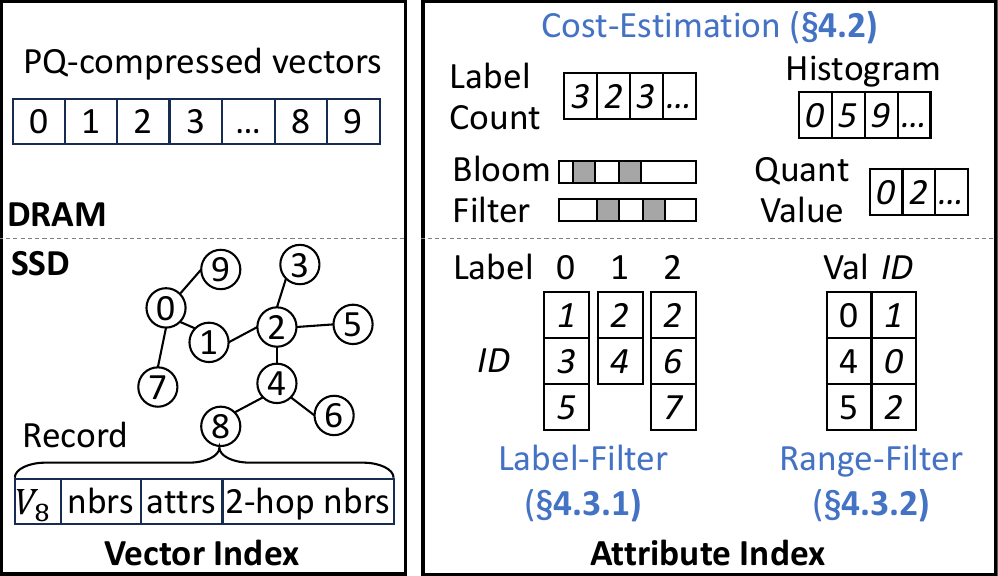}
\end{center}
\caption{\label{fig:overview} \sysname overview.
}
\end{figure}

We design and implement \sysname, a filtered ANNS system on SSD.
To build a system atop \specfilter, \sysname tackles two main design challenges:

\parahead{C1: False-positive-aware cost estimation.}
In-memory filtered ANNS systems~\cite{vldb25unify} directly use query selectivity to estimate query costs and choose filtering mechanisms (e.g., pre-filtering for low selectivity).
However, in the \specfilter scenario, this simple approach is insufficient:
First, it should consider the extra search costs caused by false positives.
Second, it should consider both SSD I/O and compute overheads simultaneously, which is more complex than in-memory scenarios.

\parahead{C2: Efficient data structures for attributes.}
To make \c{is\_member\_approx} work, the probabilistic data structures of attributes must meet three requirements: low SSD I/O, a low false-positive rate, and a small memory footprint.
Designing such data structures to support common attribute constraints (e.g., labels and ranges) is under-exploited.

\subsection{Overview}

Figure~\ref{fig:overview} shows the overview of \sysname.

\parahead{Index layout.}
\sysname consists of two parts: the vector index and the attribute index.

The vector index generally follows the graph layout described in \S\ref{sec:bg:ondisk}, which stores PQ-compressed vectors in memory and the main graph on the SSD.
However, it differs in two key aspects:
First, besides the full vector and its neighbors, each record also stores the vector's attributes, which are used for verification and post-filtering.
Second, inspired by ACORN~\cite{sigmod24acorn}, each record also stores a subset of the vector's 2-hop neighbors (neighbors of its neighbors).
Specifically, we randomly select a subset of 2-hop neighbors, making their count 10--20\x the number of direct neighbors (e.g., 96 direct neighbors and $\sim$1000 2-hop neighbors).
This design improves graph connectivity during speculative in-filtering.
\sysname only reads these extra 2-hop neighbors during in-filtering, and ignores them during pre-filtering and post-filtering.

An attribute index is built only if the attribute's filtering can be accelerated by an index.
The main index structures (e.g., inverted indexes for labels) are stored on the SSD and are scanned during pre-filtering.
In memory, \sysname maintains two types of data structures.
First, it uses memory-efficient probabilistic structures (e.g., Bloom filters and quantized values) to support the fast \c{is\_member\_approx} function.
Second, it keeps statistical summaries (e.g., histograms and label counts) for query cost estimation.
Currently, \sysname provides efficient index designs for common attribute constraints (such as labels and ranges).
However, it can be easily extended to support new, user-defined constraints.

As a result, an indexed attribute is duplicated: one copy is stored column-wise in the attribute index for scan during pre-filtering, 
while the other is stored row-wise in the vector index records for verification.

\parahead{Interfaces.} A query in \sysname takes standard vector search parameters (e.g., the query vector and $k$), plus two new inputs: the query's attributes and a \c{Selector}.
\sysname supports multiple attributes per vector, organized as a key-value (KV) map (e.g., key \c{0} mapped to a range \c{[l,r)}).

Users define filtering rules using \c{Selector} objects, which implement \c{is\_member} and \c{is\_member\_approx}.
To support query cost estimation, the \c{Selector} also exposes functions to estimate query selectivity and false-positive rate.

To optimize performance, the \c{Selector} also provides a batched interface called \c{pre\_filter\_approx}.
Depending on the implementation, this function can either evaluate base vectors one by one using \c{is\_member\_approx}, or directly perform a batched scan over an SSD attribute index to quickly return a superset of valid vector IDs.
\sysname leverages this interface in two ways:
First, for speculative pre-filtering, this function is used to get the entire valid vector superset.
Second, for speculative in-filtering, \sysname calls this interface to perform a partial attribute scan.
For example, by pre-scanning the index for rare labels, it can filter out some invalid vectors before graph traversal.
This increases the accuracy of \c{is\_member\_approx} during search.

Multiple \c{Selector}s can be combined using Boolean logic.
For example, an \c{AndSelector} takes several \c{Selector}s and applies the \c{AND} operation to their results.
Users can also fuse multiple \c{Selector}s into a single \c{Selector} for optimization.
\sysname provides built-in \c{Selector}s for common constraints (e.g., label and range filtering) and standard logical operators (\c{AND}/\c{OR}) (\S\ref{sec:impl:attr}).
Users can easily extend this interface to support custom constraint types.

\parahead{Query processing.} 
For each query, \sysname first executes cost estimation (\S\ref{sec:impl:cost}) to select the optimal execution strategy: speculative pre-filtering, speculative in-filtering, or post-filtering.
This decision balances SSD I/O overhead against memory compute costs.
Next, \sysname executes the search using the chosen method (detailed in \S\ref{sec:design}).
If pre- or post-filtering is chosen, \sysname traverses the standard graph, safely ignoring the 2-hop neighbors.

If speculative in-filtering is chosen, \sysname leverages 2-hop neighbors to maintain graph connectivity.
Specifically, at each search step, the algorithm scans both the direct and 2-hop neighbors to collect up to $R$ nodes that satisfy \c{is\_member\_approx} (i.e., possibly valid neighbors).
If it finds fewer than $R$ such nodes, it fills the remaining slots with invalid direct neighbors. 
During graph traversal, the algorithm prefers to explore possibly valid vectors first, even if some invalid vectors are geographically closer to the query.
The search terminates when the top-$L$ valid vectors are exactly verified, and no closer valid neighbors can be found.

\subsection{Cost Estimation}
\label{sec:impl:cost}

Before executing a query, \sysname uses an analytical cost model to select the optimal filtering mechanism. 
The decision relies on estimating the expected SSD I/O (pages read) and computation (distance comparisons) for each strategy.
The main challenge in this estimation is quantifying the impact of false positives introduced by speculative filtering.

\parahead{Key principles.}
To estimate the search overhead, we model the required size of the candidate pool to yield $L$ final valid vectors.
Assuming valid vectors are uniformly distributed in the dataset~\cite{sigmod26fannsbenchmark}, we apply two scaling principles:
\begin{enumerate}[itemsep=2pt,topsep=2pt,parsep=0pt,leftmargin=12pt]
    \item \emph{Selectivity scaling:} If the query selectivity is $s$, the search must explore $L/s$ vectors to find $L$ valid ones.
    \item \emph{Precision scaling:} If \c{is\_member\_approx} has a false-positive rate of (1 - $p$), namely, the precision is $p$, false positives take up slots in the candidate pool. Thus, the required pool size scales to $L/p$.
\end{enumerate}

Importantly, as discussed in \S\ref{sec:design}, false positives are not always pure overhead.
During speculative in-filtering with low selectivity, these false positives act as "bridge" edges that maintain graph connectivity. 
Because the algorithm must traverse these bridge nodes anyway to avoid early termination, the overhead of these specific false positives can be ignored in the cost model.

\begin{table}[t]
    \centering
    \begin{tabular}{l l l} 
        \toprule
        \textbf{Mechanism} & \textbf{Est. I/O (Pages)} & \textbf{Est. Compute} \\ 
        \midrule
        Pre-filtering & $X_{pre} + \frac{L}{p_{pre}} \times S_r$ & $\frac{sN}{p_{pre}}$ \\ 
        In-filtering (Low $s$) & $X_{in} + \frac{L}{s} \frac{R}{R_d} \times S_d$ & $(\frac{L}{s} \frac{R}{R_d} + \gamma \frac{L}{s}) \times R$ \\ 
        In-filtering (High $s$) & $X_{in} + \frac{L}{p_{in}} \times S_d$ & $\frac{L}{p_{in}} \times (R + \gamma R_d)$ \\ 
        Post-filtering & $\frac{L}{s} \times S_r$ & $\frac{L}{s} \times R$ \\ 
        \bottomrule
    \end{tabular}
    \caption{Cost estimation for different filtering mechanisms. 
    $N$: total base vectors. $s$: estimated query selectivity. $p_{pre}, p_{in}$: precision of pre/in-filtering (false-positive rate is $1-p$). 
    $X_{pre}, X_{in}$: pages read for initial batched attribute index scans (\c{pre\_filter\_approx}, \S\ref{sec:impl:attr}). 
    $R, R_d$: out-degree in the standard graph and the graph with 2-hop neighbors. 
    $S_r, S_d$: record size in the standard graph and the graph with 2-hop neighbors.
    $\gamma$: the relative compute cost of \c{is\_member\_approx} compared to a distance computation (we set it to 0.05 by default).}
    \label{tab:cost-estimation}
\end{table}

\parahead{Formulating the costs.}
Table~\ref{tab:cost-estimation} summarizes the estimated I/O and compute costs for the different mechanisms.

For \emph{speculative in-filtering}, the behavior has two cases depending on the selectivity.
On average, each vector in the densified graph contains $sR_d/p_{in}$ neighbors that satisfy \c{is\_member\_approx}.
When $sR_d/p_{in} \le R$ (i.e., low selectivity), the false positives serve entirely as bridge edges without introducing extra traversal overhead. 
In this case, the process is equivalent to a standard graph traversal with an effective candidate pool length of $L/s \times R/R_d$.
Conversely, when $sR_d/p_{in} > R$ (i.e., high selectivity), the false positives introduce actual overhead. The pool size must be scaled up to $L/p_{in}$ to guarantee $L$ truly valid results.

For \emph{pre-filtering}, the compute cost is dominated by the $sN/p_{pre}$ distance comparisons on PQ-compressed vectors in memory (the re-ranking compute cost is negligible).
For \emph{post-filtering}, no index scan is required, and the costs strictly follow the selectivity scaling principle.

\sysname calculates the estimated cost using a weighted linear model: $\text{Total Cost} = \alpha \times \text{Est.\ I/O} + \beta \times \text{Est.\ Compute}$, where $\alpha$ and $\beta$ are configurable weights. By default, we set $\alpha = 10$ and $\beta = 1$ to reflect the high penalty of SSD accesses.

\subsection{\SpecFilter for Attribute Constraints}
\label{sec:impl:attr}

In this section, we detail the design of \c{Selector}s for common attribute constraints.
To fully support \sysname's cost estimation and search process, each \c{Selector} must provide:
(1) \c{is\_member\_approx} and batched \c{pre\_filter\_approx} interfaces for superset filtering.
(2) Functions to estimate the filter's precision $p$ (or false-positive rate $1-p$) and query selectivity $s$ (as required in \S\ref{sec:impl:cost}).

\subsubsection{Label Filtering}
In label filtering, each vector is associated with a set of categorical labels.
A query constraint is typically a Boolean expression of labels (e.g., \c{label1 OR label2})~\cite{www23filtered}.
To execute these, we provide \c{LabelOrSelector} and \c{LabelAndSelector}.
A \c{LabelOrSelector} checks if a vector contains \emph{at least one} label from the queried labels, while a \c{LabelAndSelector} checks if it contains \emph{all} of them.
For complex Boolean queries, they can be further composed using general \c{AndSelector}s and \c{OrSelector}s.

\parahead{Index structure.}
On the SSD, labels are stored as inverted indexes: for each label, the IDs of vectors containing it are stored contiguously in ascending order.
In memory, we store each label's offset and total count. This supports fast SSD lookups and selectivity estimation with a minimal memory footprint, as the number of unique labels is typically much smaller than the dataset size.
Additionally, we maintain a lightweight, in-memory Bloom filter for each vector to enable fast, probabilistic membership checks.

\parahead{Implementation of \c{is\_member\_approx}.}
There are two baseline methods for approximate label filtering.
The first is to read the inverted indexes of all queried labels from the SSD and merge their vector IDs. This is accurate but causes severe I/O overhead for high-selectivity (frequent) labels.
The second is to rely entirely on the in-memory Bloom filters. This requires zero I/O but can become inaccurate due to hash collisions, especially for vectors with many labels.

To balance accuracy and I/O, we take a hybrid approach.
Before the graph traversal of speculative in-filtering, we only fetch the inverted indexes of \emph{low-selectivity} (rare) labels from the SSD.
We merge these fetched IDs in memory into a single list (using an intersection for \c{LabelAndSelector}, or a union for \c{LabelOrSelector}).
For the remaining \emph{high-selectivity} (frequent) labels, we fall back to the Bloom filters.

Thus, when \c{is\_member\_approx} evaluates a base vector ID during the search, it first performs a binary search to check if the ID exists in this pre-merged target list. If the ID is not in the merged rare-label list, we query the Bloom Filter for \c{LabelOrSelector}, or return \c{false} for \c{LabelAndSelector}.
This hybrid design reduces Bloom filter collisions while keeping SSD I/O strictly bounded.

For speculative pre-filtering, since it is only chosen for queries with overall low selectivity, checking the Bloom filter for every vector in the dataset would be slower than simply scanning the inverted indexes. Thus, its batched \c{pre\_filter\_approx} directly reads the required inverted indexes from the SSD and merges the IDs to generate the superset.
For \c{LabelAndSelector}, we further accelerate this process by skipping the scans for high-selectivity (frequent) labels entirely. 
Because intersecting only the low-selectivity (rare) labels already guarantees a sufficiently small superset, the high-selectivity constraints can be deferred to the final exact verification phase.

\parahead{Selectivity and precision estimation.}
We estimate query selectivity using the stored label counts. Assuming label independence, both the intersection selectivity of an \c{AND} query and the union selectivity of an \c{OR} query are easily computed.
To estimate precision, we first estimate the number of true positives using this selectivity.
Next, we estimate the number of false positives based on the Bloom filter's mathematical false-positive rate.
The final estimated precision $p$ is the ratio of true positives to the total expected positives.

\subsubsection{Range Filtering}
In range filtering, the attribute for each dataset vector is a continuous value (e.g., an integer), and the query specifies a target range $[l, r)$.

\parahead{Index structure.}
On the SSD, we store the attributes as a flat array of \c{<vector\_ID, value>} pairs, sorted by value.
This physical layout allows us to efficiently execute range queries by scanning a contiguous chunk of the array, maximizing sequential SSD reads. (Other structures, like B$^+$-trees, could also be easily integrated).

In memory, we maintain two distinct data structures with different granularities for different purposes.
For fast per-vector \c{is\_member\_approx} checks, we compress the value into a compact, 1-byte integer (256 buckets) for each vector, alongside storing the 256 global bucket boundaries.

For cost estimation, we maintain a separate, more fine-grained summary of the global value distribution. Specifically, this summary is a compact array of 1000 values representing the approximate quantiles of the dataset (e.g., the 0.1-th, ..., 99.9-th percentile values).
The key reason a 1000-quantile summary is sufficient for estimation is that our cost model does not require highly sensitive selectivity values for range queries; pre-filtering is chosen as the optimal strategy across the entire low-selectivity case.

\parahead{Implementation of \c{is\_member\_approx}.}
For speculative in-filtering, \c{is\_member\_approx} simply checks the target vector's 1-byte bucket ID in memory.
If the bucket's value range overlaps with the query range $[l, r)$, it returns \c{true}. 

For speculative pre-filtering, the batched interface queries the exact SSD index: it uses the in-memory bucket boundaries to quickly locate the starting offset, and then performs a sequential SSD read to fetch the exact valid vector IDs.

\parahead{Selectivity and precision estimation.}
We use the 1000-quantile summary to directly estimate the query selectivity.
To estimate precision, we use the same summary to estimate the number of true positives.
For the total number of positives (which includes false positives), we use the coarse-grained 256-bucket boundaries. 
Specifically, we count how many coarse buckets overlap with the query range. The estimated precision is the ratio of the two estimates.

\subsubsection{Boolean Combination}
Complex queries are handled by combining simple \c{Selector}s.
An \c{AndSelector} requires the vector to satisfy \emph{all} child constraints, while an \c{OrSelector} requires it to satisfy \emph{at least one}.

\parahead{Implementation of \c{is\_member\_approx}.}
For in-filtering, we evaluate all child branches by default, as the in-memory \c{is\_member\_approx} checks are highly optimized.

For pre-filtering, we optimize \c{AndSelector}s using early termination:
If a complex query contains a low-selectivity condition \c{AND} a high-selectivity condition across different attribute types, evaluating both on the SSD is wasteful.
Instead, the \c{AndSelector} skips the SSD scan for the high-selectivity branch. 
It only scans the low-selectivity index to generate a safe superset, deferring the high-selectivity condition to the final exact verification phase.

\parahead{Selectivity and precision estimation.}
Assuming that constraints are independent~\cite{sigmod79costestimation}:
For an \c{AndSelector}, the combined selectivity is the product of its children's selectivities.
For an \c{OrSelector}, it is the union of its children's selectivities.
Precision estimation follows a similar logic.
The combined precision of an \c{AndSelector} is the product of the individual precisions.
For an \c{OrSelector}, the combined precision is the union selectivity of the true positives divided by the union selectivity of all returned positives.

%% file: section/evaluation.tex
\section{Evaluation}
\label{sec:evaluation}

We evaluate \sysname to answer the following questions:
\begin{itemize}[itemsep=2pt,topsep=2pt,parsep=0pt,leftmargin=12pt]
    \item How does \sysname perform against state-of-the-art systems on million-scale datasets with label filtering? (\S\ref{sec:eval:overall})
    \item How does \sysname scale to 100-million-scale datasets when handling complex attribute constraints? (\S\ref{sec:eval:100m-scale})
    \item How effective are \sysname's core techniques, particularly regarding the accuracy of its cost model, the rate and impact of false-positive exploration, and the memory efficiency of its probabilistic filters? (\S\ref{sec:eval:in-depth})
\end{itemize}

\subsection{Experimental Setup}
\label{sec:setup}

\parahead{Basic configuration.} 
We conduct all experiments on a single server with the following specifications:
\begin{itemize}[itemsep=2pt,topsep=2pt,parsep=0pt,leftmargin=12pt]
    \item \b{CPU}: 2\x 28-core Intel Xeon Gold 6330 @ 2.00GHz;
    \item \b{RAM}: 512GB (16\x 32GB DDR4 2933MT/s);
    \item \b{SSD}: 1\x Samsung PM9A3 3.84TB;
    \item \b{OS}: Ubuntu 22.04 LTS with Linux kernel 5.15.0.
\end{itemize}

\parahead{Datasets.}
We use three real-world datasets:
\begin{enumerate}[itemsep=2pt,topsep=2pt,parsep=0pt,leftmargin=12pt]
    \item \emph{YFCC10M} contains image embeddings and labels.
    It consists of 10 million 192-dimensional \c{uint8} base vectors and 100,000 query vectors. Each base vector is associated with 1--1517 labels (10.8 on average), while each query has 1--2 labels (1.38 on average).
    The number of possible labels is 200,386.
    This dataset is identical to the dataset used in the BigANN benchmark~\cite{nips23bigannbenchmark}.
    
    \item \emph{YT5M} (YouTube-8M)\footnote{YouTube-8M shrinks to $\sim$5 million vectors after updates, so we call it YT5M.} contains video embeddings and labels~\cite{arxiv16youtube8M}. We use the training subset, splitting it into 5 million 1024-dimensional \c{float} base vectors and 10,000 query vectors.
    Base vectors have 1--23 labels (3.01 on average), and queries have 1--16 labels (3.05 on average).
    The number of possible labels is 3,862.
    
    \item \emph{LAION100M} contains image embeddings and metadata~\cite{arxiv21laion400m}.
    We extract the first 100 million vectors from LAION400M as base vectors, and an additional 10,000 vectors as queries.
    We extract nouns, verbs, adjectives, and adverbs from each image's text description as labels.
    Base vectors have 1--405 labels (5.69 on average), and queries have 1--10 labels (5.26 on average).
    The number of possible labels is 82,558.
    We also use the image width for range filtering.
    For queries, we generate $[l, r)$ intervals resulting in selectivities ranging from 0.001\% to 50\% (median: 15.6\%).
\end{enumerate}

\parahead{Workloads.}
We construct five types of workloads:
\begin{enumerate}[itemsep=2pt,topsep=2pt,parsep=0pt,leftmargin=12pt]
    \item \emph{Label} requires the single query label to be present in the returned vectors' labels. 
    Evaluated on LAION100M, to compare \sysname against Filtered-DiskANN~\cite{www23filtered}.
    
    \item \emph{LabelAnd} requires all query labels to be a subset of the returned vectors' labels.
    Evaluated on YFCC10M, following the BigANN benchmark~\cite{nips23bigannbenchmark}.
    
    \item \emph{LabelOr} requires the query labels to intersect with the returned vectors' labels (i.e., at least one match).
    Evaluated on YouTube8M and LAION100M.
    
    \item \emph{Range} requires a specific numerical attribute (image width) of the returned vectors to fall within a query-provided interval $[l, r)$. 
    Evaluated on LAION100M.
    
    \item \emph{Hybrid}: The returned vectors must satisfy either the LabelOr or the Range condition (a union of their results).
    Evaluated on LAION100M.
\end{enumerate}

\parahead{Compared systems.}
We compare \sysname with representative filtered ANNS systems.

\begin{enumerate}[itemsep=2pt,topsep=2pt,parsep=0pt,leftmargin=12pt]
\item \emph{PipeANN-BaseFilter}: 
An on-SSD graph-based ANNS system built atop PipeANN~\cite{osdi25pipeann}.
We implement its filtering functionality using a heuristic routing strategy: 
Queries with less than 1\% selectivity use accurate pre-filtering, while others fall back to post-filtering.
This 1\% threshold is chosen based on Figure~\ref{fig:sel-tput}, where pre-filtering and post-filtering show similar performance.

\item \emph{Milvus}~\cite{sigmod21milvus}: A popular vector database that employs pre-filtering.
We deploy Milvus v2.6.11 standalone in a local container.
We evaluate both its HNSW (\emph{Milvus-HNSW}) and IVFPQ (\emph{Milvus-IVF}) indexes.

\item \emph{Filtered-DiskANN}: An on-SSD graph-based ANNS system utilizing in-filtering.
It stores all the labels in memory and the graph index on the SSD.
Because it only supports single-label queries, we restrict its evaluation to the one-label workload on LAION100M.
\end{enumerate}

\sysname and PipeANN-BaseFilter share the same codebase, using PipeANN's PipeSearch algorithm for low-latency graph index traversal.
They also share the same graph index layout, storing attributes and the index together on the SSD.
They construct the index using the unmodified Vamana~\cite{nips19diskann} algorithm.
For \sysname, $R_d$ is selected to make each record with dense neighbors occupy one more 4KB page than the original record.
Take the LAION100M for example. Each original record occupies 4056 bytes ($\sim$4KB), and each record with dense neighbors occupies 8068 bytes ($\sim$8KB).
In contrast, Filtered-DiskANN builds a Filtered-Vamana index, which explicitly connects vectors sharing the same labels.
We find that using the same $R$ for Filtered-DiskANN and \sysname leads to poor connectivity and thus harms recall.
Therefore, we configure FilteredDiskANN with a larger $R$ (e.g., $R=256$ for LAION100M).
For Milvus-IVF, we set the number of clusters ($nlist$) to $\sim$$\sqrt{N}$ and PQ bytes to \c{dims/2}, following official recommendations~\cite{web2026milvusivfpq}.
Detailed index build parameters are summarized in Table~\ref{tab:index-build-param}.

\parahead{Metrics.}
We evaluate the latency and throughput at different recalls.
We search the top-10 nearest neighbors to get the recall (i.e., recall10@10).
We mainly compare the systems at 0.9 recall, as recommended by the BigANN~\cite{nips21bigannbenchmark,nips23bigannbenchmark} benchmark.
For latency, we use 1 search thread.
For throughput, we use enough search threads to saturate the SSD.

\begin{table}[t]
    \centering
    \setlength{\tabcolsep}{4pt}
    \begin{tabular}{l cccc cc cc} 
        \toprule
         & \multicolumn{4}{c}{\textbf{Vamana}} & \multicolumn{2}{c}{\textbf{HNSW}} & \multicolumn{2}{c}{\textbf{IVFPQ}} \\ 
        \cmidrule(lr){2-5} \cmidrule(lr){6-7} \cmidrule(lr){8-9}
        \textbf{Dataset} & $R$ & $R_d$ & $L$ & $B$ & $M$ & $efc$ & $nlist$ & $B$ \\ 
        \midrule
        YFCC10M   & 48 & 1500 & 72 & 64  & 24 & 200 & 4096 & 96  \\ 
        YT5M & 48 & 1800 & 72 & 128 & 24 & 200 & 4096 & 512 \\ 
        \bottomrule
    \end{tabular}

    \vspace{8pt} 
    
    \begin{tabular}{l cccc ccc} 
        \toprule
         & \multicolumn{4}{c}{\textbf{Vamana}} & \multicolumn{3}{c}{\textbf{Filtered-Vamana}} \\ 
        \cmidrule(lr){2-5} \cmidrule(lr){6-8} 
        \textbf{Dataset} & $R$ & $R_d$ & $L$ & $B$ & $R$ & $L$ & $B$ \\  
        \midrule
        LAION100M & 96 & 1100 & 128 & 64 & 256 & 256 & 64 \\
        \bottomrule
    \end{tabular}
    
    \caption{Index build parameters.
    We omit Milvus-HNSW and Milvus-IVF on LAION100M due to their suboptimal performance on the two smaller datasets.
    $L$ and $efc$: candidate pool length during build;
    $B$: PQ bytes per vector;
    $R$: number of edges per vector (for Vamana);
    $M$: number of edges in HNSW layers > 0 ($2M$ edges in layer 0);
    $R_d$: number of edges per vector with 2-hop neighbors;
    $nlist$: number of clusters.
    }
    \label{tab:index-build-param}
\end{table}

\subsection{Overall Performance}
\label{sec:eval:overall}

\begin{figure}[t]
\begin{center}
\includegraphics[width=\linewidth]{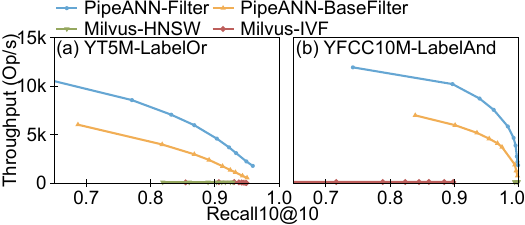}
\end{center}
\caption{Search throughput on YT5M and YFCC10M.}
\label{fig:eval:overall-tput}
\end{figure}

In this section, we evaluate \sysname on two label-filtering datasets: YT5M and YFCC10M.
YT5M evaluates label \c{OR} conditions, while YFCC10M evaluates label \c{AND} conditions.
We compare \sysname against PipeANN-BaseFilter and Milvus.

\parahead{Throughput.}
Figure~\ref{fig:eval:overall-tput} shows the search throughput.
From the figure, we make the following observations:

(1) \sysname achieves higher throughput than PipeANN-BaseFilter.
At 0.9 recall, \sysname reaches 1.91\x and 1.71\x the throughput of PipeANN-BaseFilter on YT5M and YFCC10M, respectively.
This improvement comes from \sysname's efficient in-filtering powered by \specfilter.
For instance, at 0.9 recall, it processes 77.6\% and 50.4\% of queries using speculative in-filtering on these two datasets.
In contrast, PipeANN-BaseFilter relies entirely on pre- and post-filtering, both of which incur higher I/O overhead than in-filtering for queries with moderate selectivity and recall targets.

At higher recall targets, this throughput advantage grows significantly:
\sysname reaches 3.95\x and 2.03\x the throughput of PipeANN-BaseFilter on YT5M (at 0.95 recall) and YFCC10M (at 0.99 recall).
This gap stems from the rapidly increasing cost of post-filtering.
While both in-filtering and post-filtering require linearly more I/O to achieve higher accuracy (with a larger $L$), post-filtering's cost grows at a much steeper rate (as shown in Table~\ref{tab:cost-estimation}).

Furthermore, the performance gain is larger on YT5M than on YFCC10M.
This occurs because PipeANN-BaseFilter treats YT5M as a post-filtering-intensive workload (78.8\% post-filtering), whereas YFCC10M is pre-filtering-intensive (only 29.7\% post-filtering).
This highlights the efficiency of speculative in-filtering over post-filtering.

(2) Milvus suffers from low throughput due to its strict pre-filtering policy.
Its throughput peaks at under 150 ops/s across all recall targets.
While Milvus performs well in standard vector search, its filtered ANNS throughput is severely bottlenecked by heavy attribute scans during pre-filtering.
Even with inverted index acceleration, these scans strictly dominate the overall execution time.

\begin{figure}[t]
\begin{center}
\includegraphics[width=\linewidth]{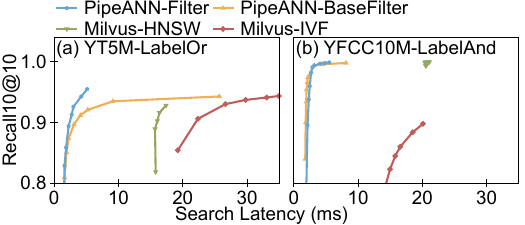}
\end{center}
\caption{Search latency on YT5M and YFCC10M.}
\label{fig:eval:overall-lat}
\end{figure}

\parahead{Latency.}
Figure~\ref{fig:eval:overall-lat} shows the latency results.
From the figure, we make the following observations:

(1) \sysname maintains competitive latency compared to PipeANN-BaseFilter.
At 0.9 recall, \sysname's average search latency is 72.7\% and 1.11\x of PipeANN-BaseFilter's latency on YT5M and YFCC10M, respectively.
On YT5M, \sysname reduces latency because it uses in-filtering for moderate-selectivity queries, which creates shorter search paths than PipeANN-BaseFilter's post-filtering.
Conversely, on YFCC10M, \sysname shifts 20.7\% of the total queries from PipeANN-BaseFilter's pre-filtering to in-filtering.
While pre-filtering has poor throughput due to I/O overheads, its latency can be lower under light SSD loads, because its attribute index and vector reads are highly parallelized. In contrast, speculative in-filtering relies on sequential graph traversal.

(2) Milvus exhibits high latency.
While it reaches a similar recall to \sysname, its average latency easily exceeds 10ms, again bottlenecked by attribute index scans.
Due to this poor performance, we exclude Milvus from the remaining 100M-scale dataset evaluations.

\subsection{100M-Scale Dataset}
\label{sec:eval:100m-scale}
\begin{figure}[t]
\begin{center}
\includegraphics[width=\linewidth]{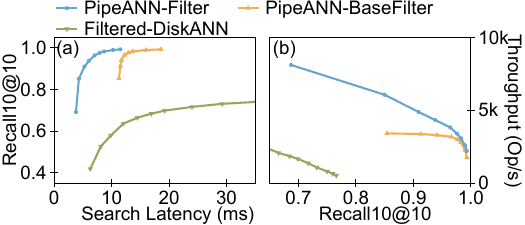}
\end{center}
\caption{Search latency and throughput of single-label filtering. Dataset: LAION100M.}
\label{fig:eval:single-label}
\end{figure}

\parahead{Single-label performance.} 
Here, we evaluate single-label filtering on LAION100M against PipeANN-BaseFilter and Filtered-DiskANN.
Figure~\ref{fig:eval:single-label} shows the results.
We find that \sysname's performance gains over PipeANN-BaseFilter mirror our previous observations in \S\ref{sec:eval:overall}:
At 0.9 recall, \sysname achieves 1.44\x the throughput and reduces average latency to 45.5\% of PipeANN-BaseFilter.
At the higher 0.99 recall target, these improvements drop slightly to 1.14\x throughput and 55.1\% latency.
The throughput trend differs slightly from \S\ref{sec:eval:overall} because this workload is heavily pre-filtering-intensive (only 8.49\% post-filtering, even less than YFCC10M).
While post-filtering I/O still grows faster than \sysname's in-filtering, the absolute number of post-filtering queries is so small that the average throughput drop is less severe.
Moreover, to achieve high recall, in-filtering requires heavy sequential graph traversal, making it more expensive than pre-filtering (which only needs to read more vectors for re-ranking).
As a result, \sysname selects in-filtering less frequently (e.g., 32.8\% at 0.9 recall vs. 18.6\% at 0.99 recall) at high recall targets, causing its latency to converge with PipeANN-BaseFilter.

Filtered-DiskANN performs poorly, failing to reach a recall above 0.8.
Even though it uses a larger out-degree than other graph indexes, its pruning process still disconnects some vectors sharing the same label.
This breaks graph connectivity during in-filtering, severely limiting recall.
While \sysname also relies on speculative in-filtering for many queries (e.g., 32.8\% at 0.9 recall), it naturally preserves connectivity using the "bridge" nodes provided by \specfilter.
This allows it to achieve high recall with fewer neighbors.

\begin{figure}[t]
\begin{center}
\includegraphics[width=\linewidth]{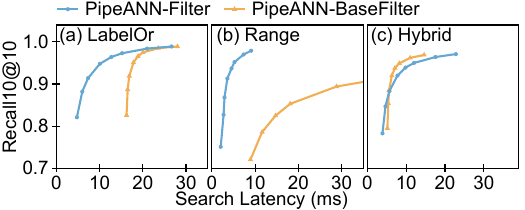}
\end{center}
\caption{Search latency on LAION100M.}
\label{fig:eval:latency-laion100m}
\end{figure}

\begin{figure}[t]
\begin{center}
\includegraphics[width=\linewidth]{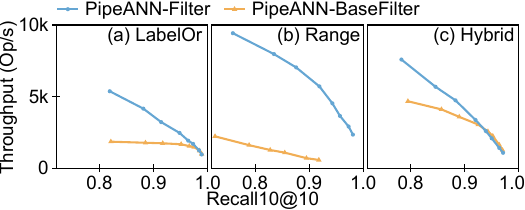}
\end{center}
\caption{Search throughput on LAION100M.}
\label{fig:eval:tput-laion100m}
\end{figure}

\parahead{Other workloads.}
Figures~\ref{fig:eval:latency-laion100m} and~\ref{fig:eval:tput-laion100m} show the latency and throughput on other LAION100M workloads.
Generally, the performance trends match those of previous evaluations.

The performance gain of \sysname is most significant in the range-filtering ANNS workload:
At 0.9 recall, \sysname achieves 9.78\x the throughput and reduces latency to 12.5\% of PipeANN-BaseFilter.
For PipeANN-BaseFilter, this workload is post-filtering-intensive (75.6\%), where PipeANN-BaseFilter fails to reach a high recall within a 10ms latency scale.
This shows that post-filtering sometimes struggles to find enough valid nearby vectors under tight range constraints.
In contrast, \sysname's speculative in-filtering maintains graph connectivity, delivering both superior recall and throughput.

\subsection{In-Depth Analysis}
\label{sec:eval:in-depth}

In this section, we examine the cost estimation accuracy, the false-positive rate of speculative in-filtering, and the memory usage of \sysname's probabilistic filters.

\parahead{Cost estimation accuracy.}
We find that the cost estimation for speculative pre-filtering is highly accurate, as its index scan and brute-force search costs are highly predictable.
Furthermore, thanks to low query selectivity for all labels, speculative pre-filtering translates into exact attribute index scans in our workloads.
Thus, we focus our discussion on in-filtering and post-filtering, specifically comparing their estimated I/O versus actual I/O.
We omit the compute cost here because it is proportional to the I/O volume during both graph traversal and attribute index scans.

\begin{figure}[t]
\begin{center}
\includegraphics[width=\linewidth]{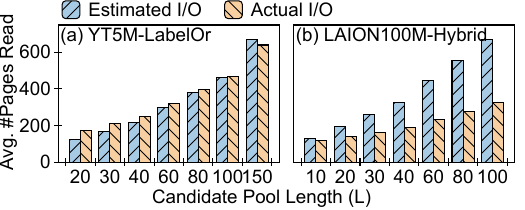}
\end{center}
\caption{I/O estimation for speculative in-filtering.}
\label{fig:eval:infilter-est}
\end{figure}

Figure~\ref{fig:eval:infilter-est} shows the in-filtering results for two representative workloads, YT5M and LAION100M (other workloads show similar trends).
For YT5M, the I/O estimation is quite accurate (0.74\x to 1.04\x of actual I/O). However, it is overly conservative for LAION100M (estimating up to 2.05\x the actual I/O).
This overestimation happens because of \sysname's early termination mechanism.
The search stops as soon as it finds $L$ valid vectors that cannot be updated by closer neighbors.
In contrast, the cost model conservatively assumes a search budget of $L/s$ (where $s$ is selectivity).
Although they are statistically equivalent when the valid vectors are uniformly distributed, they are not when the valid vectors show different distributions (e.g., clustered).

\begin{figure}[t]
\begin{center}
\includegraphics[width=\linewidth]{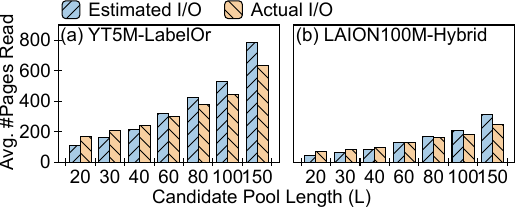}
\end{center}
\caption{I/O estimation for post-filtering.}
\label{fig:eval:postfilter-est}
\end{figure}

Figure~\ref{fig:eval:postfilter-est} shows the results for post-filtering.
From the figure, we observe that the estimated I/O initially underestimates the actual I/O, but then grows to overestimate it.
The initial underestimation occurs because our model omits the $\log(N)$ baseline cost of traversing the graph toward the target region~\cite{pr1980rng}, assuming high search $L$ values will dominate the cost.
At larger $L$ values, it overestimates the cost due to the same early termination effect seen in in-filtering.

In conclusion, developing a data-distribution-aware cost model that accounts for early termination is an interesting direction for future work.

\parahead{False-positive exploration rate.}
\Specfilter intentionally explores false-positive vectors.
In our setup, this occurs almost entirely during in-filtering, as pre-filtering executes precise index scans.
Thus, we focus on the in-filtering false-positive rate.
Overall, the false-positive rate ranges from 0.8\% to 69.4\%, averaging 23.6\% with a median of 17.1\%.
This confirms that \sysname's lightweight filters remain highly effective.
In the worst-case scenario (LAION100M-1Label), \sysname experiences an average false-positive rate of 63.5\% across high-recall targets (0.85 to 0.99).
Despite this high rate, \sysname's speculative in-filtering still outperforms the strict pre- and post-filtering of PipeANN-BaseFilter (as shown earlier in Figure~\ref{fig:eval:single-label}).
\begin{table}[t]
\centering
\setlength{\tabcolsep}{6pt}
\begin{tabular}{llcc}
\toprule
Dataset & Type & \makecell{In-Memory \\ Filter Size} & \makecell{Ratio to On-SSD \\ Attribute Index} \\
\midrule
YFCC10M   & Label & 39MB  & 3.5\% \\
YT5M      & Label & 20MB  & 28.9\% \\
LAION100M & Label & 384MB & 15.8\% \\
LAION100M & Range & 96MB  & 12.5\% \\
\bottomrule
\end{tabular}
\caption{Memory usage of \sysname's probabilistic filters.}
\label{tab:memory-usage}
\end{table}
\parahead{Memory usage.}
Table~\ref{tab:memory-usage} describes the memory usage of \sysname's probabilistic filters.
For label attributes, \sysname applies per-vector Bloom filters, consuming a fixed 4 bytes per vector.
Because the average number of labels per vector varies across datasets (ranging from 3.01 to 10.8), these Bloom filters account for 3.5\% to 28.9\% of the total attribute index size.
Generally, a larger memory footprint correlates with better filtering performance (e.g., YFCC10M vs. YT5M).
However, \sysname remains efficient even on YFCC10M, where the memory-to-SSD index ratio is merely 3.5\%.
For range attributes, \sysname quantizes values (from 4 bytes down to 1 byte).
This requires only 12.5\% of the memory used by the full SSD attribute index (which stores 8 bytes per vector for the ID and raw value).

%% file: section/related_work.tex
\section{Related Work}
\label{sec:related}

\parahead{On-SSD ANNS.}
Large-scale vector datasets (e.g., billion-scale~\cite{web2021microsoftann,vldb2020adbv}) require terabytes of storage space.
Storing vectors and indexes on the SSD is a practical approach to support vector search in a scalable and cost-effective manner.

Many existing works focus on graph-based ANNS on the SSD.
Typically, they follow a graph layout similar to DiskANN~\cite{nips19diskann}: the full graph is stored on the SSD, while PQ-compressed vectors are kept in memory to guide graph navigation.
These works primarily address two hardware characteristics of SSDs compared to DRAM: larger access granularity and higher access latency.
First, an SSD's access unit (a page) is much larger than a single graph record (typically hundreds of bytes), leading to unused space per read. 
To improve I/O efficiency, systems pack nearby records~\cite{sigmod24starling,arxiv25gorgeous} or additional neighbors~\cite{arxiv25pageann} into this unused space, thus reducing the number of I/O requests and increasing throughput.
Second, an SSD's high access latency makes the search process heavily I/O-bound. 
To mitigate this, previous works propose compute-I/O overlapping~\cite{sigmod24starling,arxiv26alayalayser,osdi25pipeann}, or aim to reduce the total number of search hops via entry point optimization~\cite{sigmod24starling,sigmod26gustann} and early stopping~\cite{sigmod20termination}.
For a comprehensive survey, we refer readers to~\cite{arxiv26diskgraphsummary}.

Other works explore cluster-based ANNS on the SSD.
These systems group vectors into clusters during index building. During a search, they only scan the clusters whose centers are closest to the query.
Unlike in-memory IVF~\cite{cvpr12ivf,eccv18ivfhnsw}, reading large clusters from the SSD is slow.
Therefore, SPANN~\cite{nips21spann,sosp23spfresh} adopts a fine-grained hierarchical clustering approach, where each cluster contains only tens of vectors (instead of the $\sqrt{N}$ recommended by Faiss). An in-memory graph is then used to index these cluster centers (e.g., a 100M-scale graph for 1 billion vectors).   
Because cluster-based indexes rely on brute-force scanning within clusters, they can also be compute-intensive. Some recent works accelerate this scanning process using GPUs~\cite{fast25fusionanns} or SmartSSDs~\cite{atc24smartann}.

Unlike the above systems, which focus on standard (unfiltered) top-$k$ vector search, \sysname addresses the I/O bottlenecks of top-$k$ search \emph{with attribute filters}.

\parahead{Filtered ANNS.}
There are two primary approaches to efficient filtered ANNS: attribute-aware vector indexing and attribute-agnostic search.

\emph{Attribute-aware vector indexes} modify the underlying graph structure. They adjust the edge selection mechanism to connect vectors that are highly likely to be accessed together under specific attribute constraints.
These works mainly target two workloads: label filtering and range filtering.
For label filtering, systems typically connect vectors that share overlapping labels~\cite{www23filtered,nips23nhq,sigmod24ung}.
Range filtering is more complex. One approach builds the index by explicitly connecting vectors with overlapping ranges~\cite{sigmod24serf}. Another approach partitions the dataset into sub-graphs based on value ranges, and then connects these sub-graphs~\cite{vldb25unify,sigmod24irangegraph}.
Because the graph structure is deeply customized for specific attributes, this approach can achieve very high performance.
However, building and maintaining separate indexes for different attributes is space-inefficient, and an index optimized for one set of attributes may perform poorly on others.

\emph{Attribute-agnostic search} relies on a single, general-purpose vector index for all queries, regardless of the attribute constraints.
It typically employs the classic mechanisms: pre-filtering, in-filtering, and post-filtering.
Due to its high generalizability and low storage overhead, modern vector databases like Faiss~\cite{douze2024faiss} and Milvus~\cite{sigmod21milvus} adopt this approach.
Notably, ACORN~\cite{sigmod24acorn} proposes densifying the standard graph index to maintain connectivity during in-filtering, an insight that inspired \sysname's on-SSD graph layout.
While highly general, attribute-agnostic search typically shows lower performance compared to attribute-aware approaches~\cite{sigmod26fannsbenchmark}.

\sysname follows the attribute-agnostic design for generalizability. 
By introducing \specfilter, \sysname overcomes the I/O bottlenecks that traditionally make pre- and in-filtering impractical on the SSD.

\parahead{Probabilistic filters.}
Probabilistic filters, such as Bloom filters~\cite{cacm70bloomfilter}, are highly compact data structures designed for fast, approximate membership testing.
Storage systems, like Log-Structured Merge (LSM) trees~\cite{web2026rocksdb}, adopt them to prevent expensive, unnecessary disk reads.
Similarly, distributed systems use them to reduce network I/O~\cite{sigcomm05hashlookup,conext09buffalo}.
Over time, many advanced probabilistic filters have emerged.
For example, the Cuckoo filter~\cite{conext14cuckoofilter} supports item deletion (unlike basic Bloom filters), and SuRF~\cite{sigmod18surf} supports both single-key lookups and range queries.

Strict filtering cannot adopt probabilistic filters because it cannot tolerate false positives during the search.
In contrast, \specfilter enables their use based on our key observation that the initial filtering need not be accurate, and that subsequent verification is lightweight.

%% file: section/conclusion.tex
\section{Conclusion}
We propose \sysname, a filtered ANNS system on SSD.
Leveraging \specfilter, \sysname mitigates the I/O bottleneck of traditional filtering mechanisms.
Evaluations show that \sysname boosts performance compared to existing filtered ANNS systems on SSD.
This work shows that design philosophies from traditional storage systems, such as leveraging probabilistic data structures to reduce disk reads, can be repurposed to accelerate modern vector storage systems.